\documentclass[pra,aps,twocolumn,
longbibliography,superscriptaddress]{revtex4-2}
\usepackage[utf8]{inputenc}
\usepackage[T1]{fontenc}
\usepackage{amsmath,amssymb,amsthm}
\usepackage{braket,bm}
\usepackage{hyperref}

\usepackage{xcolor}
\definecolor{maroon}{RGB}{100,20,20}
\definecolor{dblue}{RGB}{20,20,100}

\hypersetup{colorlinks=true,
linkcolor=dblue,
citecolor=maroon,
urlcolor=maroon}

\theoremstyle{plain}

\newtheorem{proposition}{Proposition}

\theoremstyle{definition}
\newtheorem{definition}{Definition}
\theoremstyle{remark}

\newcommand{\C}{\mathbb{C}}
\newcommand{\ot}{\otimes}
\newcommand{\HH}{\mathcal{H}}
\newcommand{\Cc}{\mathcal{C}}

\newcommand{\Tr}{\operatorname{Tr}}
\newcommand{\rank}{\operatorname{rank}}
\newcommand{\range}{\operatorname{range}}
\newcommand{\supp}{\operatorname{supp}}
\newcommand{\spa}{\operatorname{span}}
\usepackage[normalem]{ulem}

\begin{document}
\title{Positive-partial-transpose structure of completely
entangled subspaces and their complements}
\author{Gurvir Singh}
\affiliation{Department of Physical Sciences, Indian
Institute of Science Education and Research Mohali, Sector
81, SAS Nagar, Punjab 140306, India}
\affiliation{Harish-Chandra Research Institute, Chhatnag
Road, Jhunsi, Prayagraj 211019, India}
\author{Arvind}
\affiliation{Department of Physical Sciences, Indian
Institute of Science Education and Research Mohali, Sector
81, SAS Nagar, Punjab 140306, India}

\begin{abstract}
A completely entangled subspace (CES) contains no product
vectors, so every state supported on it is entangled.
Whether such a subspace can support states
with positive partial transpose (PPT) is a more subtle
question.
The normalized projector onto a CES may be
NPT even if there exist PPT states supported on the CES.
We investigate this distinction using a semidefinite
criterion that optimizes the smallest partial-transpose
eigenvalue over all supported density operators. For two-
qutrit states supported on CESs, this leads to a complete
characterization: a CES supports PPT states exactly when it
is four-dimensional and related by invertible local
transformations to the complement of an unextendible product
basis (UPB). We also establish restrictions on PPT states
whose ranges contain only vectors of Schmidt rank at least
three, ruling out their existence in $3\otimes n$,
$4\otimes4$, and $4\otimes5$, and showing that any such
state in $4\otimes6$ must have rank eight. An interesting
structure emerges when two orthogonal completely entangled
subspaces together span the full Hilbert space, which we
term \textit{complementary CESs}. Explicit constructions in
$3\otimes4$ demonstrate that both subspaces can support PPT
entanglement even when their normalized projectors are NPT.
In $4\otimes4$, we construct a complementary pair whose
normalized projectors are themselves PPT entangled. These
constructions offer an alternative route to obtaining PPT
entangled states, beyond the familiar approach based on UPB
complements.
\end{abstract}
\maketitle

\section{Introduction}
Entanglement is central to quantum communication,
computation, and
metrology~\cite{Nielsen2010,Pirandola2020,Giovannetti2011}.
While for bipartite pure states is possible to characterize
entanglement, for mixed states, however, determining
separability is
generally difficult~\cite{Horodecki2009}. The positive-
partial-transpose (PPT)
criterion~\cite{Peres1996,Horodecki1997} is a basic tool in
this setting: every separable state is PPT, but PPT
entangled states occur already in
$3\otimes3$~\cite{Horodecki1998Bound}. Such PPT entangled
states
which are also known as bound-entangled states are
undistillable.
Even though Bound entangled states are undistillable, they
are known to possess nontrivial information-theoretic
properties and applications, including the possibility of
distilling secret key~\cite{Horodecki2005Key} and their use
in quantum metrology~\cite{Toth2018,Pal2021}.
Completely entangled subspace(CES) are those subspaces
of a composite system which do not support any seperable
state~\cite{Parthasarathy2004}. While every state in a CES
is entangled PPT
there can be NPT as well PPT states within a
CES~\cite{DiVincenzo2003}.
Understanding which subspaces admit PPT entanglement is
therefore a natural part of the study of mixed-state
entanglement.

The fact that a CES supports only entangled states makes
it a natural setting for constructing PPT entangled states:
any PPT density operator supported on a CES is necessarily
entangled. A well-known example comes from orthogonal
unextendible product bases (UPBs). The orthogonal complement
of a UPB is a CES, and the normalized projector onto this
subspace is PPT entangled~\cite{Bennett1999,DiVincenzo2003,
Bravyi2004}.

The study of entangled subspaces also extends beyond the
absence of product vectors. In bipartite systems, a stronger
condition is obtained by requiring every nonzero vector in
a subspace to have Schmidt rank at least $r$. Such subspaces
are known as $r$-entangled subspaces, with ordinary CESs
corresponding to $r=2$~\cite{Cubitt2008,Lovitz2022}.
In multipartite systems, considerable attention has been
devoted to genuinely entangled subspaces (GESs), which
contain no nonzero vectors separable across any
bipartition. Consequently, every state supported on a GES
is genuinely multipartite entangled~\cite{
Demianowicz2018,sristy2019,DemianowiczAugusiak2020,
Demianowicz2022a,Demianowicz2022b,
MakutaAugusiak2025}.

The notions discussed above characterize subspaces in
terms of the kinds of pure states they exclude. The
question of PPT entanglement is somewhat different:
given a bipartite CES, can any of the mixed states
supported on it have positive partial transpose?
The normalized projector provides a natural starting
point, as illustrated by the UPB construction, but
it represents only one of the possible states supported
on the subspace. While a PPT normalized projector
establishes the existence of PPT entanglement, an
NPT normalized projector does not rule out other
PPT states supported on the same CES.
Johnston studied subspaces in which every supported state is NPT~\cite{Johnston2013}, while Sengupta et al. investigated positive operators supported on a particular maximal CES~\cite{Ritabrata2014}.
We therefore consider the full set of supported
density operators and introduce a quantity
$\mu(\Cc)$, obtained by maximizing the smallest
eigenvalue of the partial transpose over this set.
This optimization provides a semidefinite criterion
for determining whether a CES supports PPT
entangled states. In $3\otimes3$, the known classification
of rank-four
PPT entangled states~\cite{Chen2011,Skowronek2011,
ChenDjokovic2012classes} allows us to obtain a complete
characterization of PPT support for completely entangled
subspaces.

Recently, considerable attention has been devoted to the
construction of PPT entangled states with increasingly
large Schmidt number~\cite{
HuberLamiLancienMullerHermes2018,
Cariello2020,KrebsGachechiladze2024,
KrebsGachechiladze2026}.
This motivates a related question:
can a PPT state be supported on a completely entangled
subspace in which every nonzero vector has Schmidt rank
at least three? Such a subspace excludes not only
product vectors, as required of any CES, but also
all vectors of Schmidt rank two. This requirement is
stronger than asking for a PPT state of Schmidt number
at least three. While the Schmidt number constrains the
possible pure-state decompositions of a mixed state,
the condition considered here applies to every vector
in its range. We investigate whether PPT entanglement
is compatible with this stronger restriction and show
that no such PPT state exists in $3\otimes n$,
$4\otimes4$, or $4\otimes5$. In $4\otimes6$,
we show that any PPT state satisfying this condition
must have rank eight.

Another interesting possibility arises when the orthogonal
complement of a CES is itself completely entangled.
In this case, the entire Hilbert space can be decomposed
into two orthogonal completely entangled subspaces,
which we refer to as complementary CESs. This is
particularly interesting in contrast to the familiar
UPB construction, where the orthogonal complement
of the CES is spanned by product vectors.
Although complementary CESs have been studied
previously~\cite{Augusiak2011CES,Skowronek2013,
HalderAugusiak2025}, to the best of our knowledge, their
PPT-support properties have not been given due attention,
Using the semidefinite framework developed in this work,
we address precisely this question: can PPT entangled
states be supported on one or both CESs in a
complementary pair?
We establish a constraint on the simultaneous
existence of PPT states on complementary CESs,
showing that their partial transposes must necessarily
be singular. Through explicit constructions in
$3\otimes4$, we demonstrate that both subspaces
can support PPT entanglement even if their
normalized projectors are NPT. We further construct
a complementary pair in $4\otimes4$ case whose normalized
projectors are themselves PPT entangled, providing
an alternative way to constructing PPT entangled
states beyond the usual UPB-based approach.

The paper is organized as follows.
In Section~\ref{preliminaries}, we introduce the relevant
definitions and notation. In Sec.~\ref{PPT_projections},
we investigate the PPT properties of CESs and establish
a complete characterization for the two-qutrit case.
Section~\ref{sec:high-sr} examines PPT support in
subspaces with minimum Schmidt rank three.
In Sec.~\ref{sec:complementary}, we study complementary
CESs and present explicit constructions in $3\otimes4$
and $4\otimes4$. Finally, in Sec.~\ref{conclusion}
the conclusions are presented.

\section{Preliminaries}
\label{preliminaries}

Before examining the PPT properties of CESs, we briefly
recall the relevant notions of separability and entangled
subspaces and establish the notation used throughout.

Let $\mathcal H_{d_1,\ldots,d_n}
=\mathbb C^{d_1}\otimes\cdots\otimes\mathbb C^{d_n}$
denote the Hilbert space of an $n$-partite quantum system.
A pure state is fully separable if it is a tensor product
of states belonging to the individual subsystems. More
generally, a mixed state $\rho$ is fully separable if it
admits a decomposition of the form
\begin{equation}
\rho=\sum_k p_k\bigotimes_{j=1}^{n}
|a_k^{(j)}\rangle\langle a_k^{(j)}|,
\end{equation}
where $p_k\geq 0$, $\sum_k p_k=1$, and
$|a_k^{(j)}\rangle\in\mathbb C^{d_j}$
are pure states of the respective subsystems.
For bipartite systems, a state that does not admit such
a decomposition is entangled. In multipartite systems,
one can also consider biseparability, where a mixed state
is a convex combination of pure states that are product
across some bipartition, with the bipartition allowed to
vary between terms. States that are not biseparable are
called genuinely multipartite entangled (GME)
states~\cite{Horodecki2009,Demianowicz2018}.

For a bipartite system, a density operator $\rho$
is said to have positive partial transpose (PPT)
if $\rho^{T_B}\geq 0$, where $T_B$ denotes
transposition with respect to the second subsystem
in a fixed local basis. If $\rho^{T_B}$ has at
least one negative eigenvalue, the state is
called NPT.

Every bipartite pure state $\ket{\psi}$ admits
a Schmidt decomposition of the form
\[
\ket{\psi}=
\sum_{j=1}^{s}\sqrt{\lambda_j}
\ket{a_j}\ot\ket{b_j},
\]
where $\{\ket{a_j}\}$ and $\{\ket{b_j}\}$ are
orthonormal sets of local vectors, and
$\lambda_j>0$ are the Schmidt coefficients
satisfying $\sum_j\lambda_j=1$.
The number $s$ of nonzero Schmidt coefficients
is called the Schmidt rank of the state,
denoted by $\operatorname{SR}(\psi)$.
For a bipartite mixed state $\rho$, the
Schmidt number is defined as
\begin{equation}
\operatorname{SN}(\rho)=
\min_{\rho=\sum_j p_j\ket{\psi_j}\bra{\psi_j}}
\max_j\operatorname{SR}(\psi_j),
\label{eq:schmidt-number}
\end{equation}
where the minimum is taken over all pure-state
decompositions of $\rho$. A bipartite state
is separable if and only if its Schmidt number
is one. It is worth noting that the Schmidt
number characterizes the possible pure-state
decompositions of a mixed state and does not
impose a lower bound on the Schmidt rank of
every vector in its range.

\begin{definition}[Completely entangled subspace]
\label{def:CES}
A subspace $\Cc\subseteq\HH_{d_1,\ldots,d_n}$
is completely entangled if it contains no
nonzero fully product vector.
\end{definition}

Consequently, every density operator supported
on a CES is entangled. For a bipartite system
$\C^{d_A}\ot\C^{d_B}$, the maximum possible
dimension of a CES is given
by~\cite{Parthasarathy2004}
\begin{equation}
\dim\Cc_{\max}=(d_A-1)(d_B-1).
\label{eq:CESdim}
\end{equation}

A stronger condition can be imposed by requiring
the subspace to contain no vectors below a
specified Schmidt rank. A bipartite subspace
$\Cc$ is called $r$-entangled if
\begin{equation}
\operatorname{SR}(\psi)\geq r
\quad\text{for every }0\neq\psi\in\Cc.
\label{eq:rCES}
\end{equation}
The maximum dimension of such a subspace
is bounded by~\cite{Cubitt2008}
\begin{equation}
\dim\Cc\leq
(d_A-r+1)(d_B-r+1).
\label{eq:rCESdim}
\end{equation}
Thus, an ordinary bipartite CES corresponds
to the case $r=2$.

For multipartite systems, a related notion
is that of a genuinely entangled subspace
(GES), which contains no nonzero vectors
separable across any bipartition.
Every state supported on a GES is therefore
genuinely multipartite entangled.
A multipartite CES, however, need not be a
GES, since the absence of fully product
vectors does not exclude vectors that are
separable across a bipartition involving
groups of subsystems~\cite{
Demianowicz2018,DemianowiczAugusiak2020}.

For any subspace $\Cc$, we denote its
orthogonal projector by $P_\Cc$ and define
the corresponding normalized projector as
\begin{equation}
\rho_\Cc=\frac{P_\Cc}{\dim\Cc},
\label{eq:projstate}
\end{equation}
which is a density operator satisfying
$\operatorname{Tr}\rho_\Cc=1$ and
$\operatorname{supp}(\rho_\Cc)=\Cc$.

A familiar way of constructing CESs is through
unextendible product bases (UPBs). An orthogonal UPB
consists of mutually orthogonal product vectors that
do not span the entire Hilbert space, but for which
no additional product vector can be found orthogonal
to all of them. More precisely, for a collection
$\mathcal U=\{\ket{a_j}\ot\ket{b_j}\}_{j=1}^{k}$,
with $k<d_Ad_B$, this means
\begin{equation}
(\operatorname{span}\mathcal U)^\perp
\cap
\{\ket a\ot\ket b:\ket a,\ket b\neq0\}
=\varnothing.
\label{eq:upbdef}
\end{equation}
The orthogonal complement of a UPB is therefore a CES.
An interesting property of this construction is that
the normalized projector onto the complement is always
PPT. This follows because partial transposition
takes the orthogonal product projectors of the UPB
to another set of orthogonal product projectors.
The resulting state is thus PPT entangled
~\cite{Bennett1999,Bravyi2004}.

For an arbitrary CES, however, the normalized projector
need not be PPT. This does not necessarily mean that
the subspace cannot support PPT entanglement, since
other density operators supported on the same CES
may still have a positive partial transpose.
\section{PPT structure of completely entangled subspaces}
\label{PPT_projections}
We now examine the PPT properties of the states which a CES
can support. Our aim is to determine whether there exists at
least
one PPT state or whether all states supported on a subspace
are NPT.
While for UPB, the normalized projector is able to settle
this question, this is not the case for general CESs.
Therefore, in order
examine arbitrary CESs, we consider the entire set of
supported set of
supported density operators.

Let $Q$ be the $d_Ad_B\times r$ matrix whose columns form an
orthonormal basis of $\Cc$, where $r=\dim\Cc$. Every
supported density operator can be written as
$\rho=QAQ^\dagger$ with $A\geq0$ and $\Tr A=1$. We define
\begin{equation}
\mu(\Cc)=\max_{\substack{\rho\geq0,\
\Tr\rho=1\\\supp(\rho)\subseteq\Cc}}
\lambda_{\min}(\rho^{{\rm T}_B}).
\label{eq:mudef}
\end{equation}

The optimization over supported states can be carried
out directly in terms of $A$. Introducing a variable
$t$ for the smallest eigenvalue of the partial transpose
gives the semidefinite program
\begin{equation}
\begin{aligned}
\text{maximize}\quad &t\\
\text{subject to}\quad&A\geq0,\quad\Tr A=1,\\
&(QAQ^\dagger)^{T_B}-tI_{d_Ad_B}\geq0.
\end{aligned}
\label{eq:muSDP}
\end{equation}
The value of $\mu(\Cc)$ does not depend on the
choice of orthonormal basis for $\Cc$.
The corresponding dual expression is
\begin{equation}
\mu(\Cc)=\min_{\substack{W\geq0\\\Tr W=1}}
\lambda_{\max}\bigl(Q^\dagger W^{T_B}Q\bigr).
\label{eq:dual}
\end{equation}
This dual has a useful interpretation in terms of
entanglement witnesses. For $W\geq0$, the operator
$W^{T_B}$ has nonnegative expectation on every PPT
state, since
$\Tr(W^{T_B}\rho)=\Tr(W\rho^{T_B})\geq0$.
Therefore, whenever it detects an entangled state, $W^{T_B}$
is a decomposable entanglement witness.
If the dual optimum is negative, the corresponding
witness has negative expectation on every density
operator supported on $\Cc$. Indeed, for any
$\rho=QAQ^\dagger$,
\[
\Tr(W^{T_B}\rho)
\leq\lambda_{\max}(Q^\dagger W^{T_B}Q)<0.
\]
Thus, a single decomposable witness can detect
every state supported on a CES, rather than
merely its normalized projector~\cite{Augusiak2011CES}.

The interpretation of $\mu(\Cc)$ follows from the
optimization itself. Since the set of supported
density operators is compact, the maximum is
always attained. For a CES, $\mu(\Cc)<0$ implies
that every supported state is NPT, whereas
$\mu(\Cc)\geq0$ guarantees the existence of at
least one PPT entangled state. Furthermore,
$\mu(\Cc)>0$ means that a supported state has a
positive-definite partial transpose. At
$\mu(\Cc)=0$, PPT states still exist, but their
partial transposes are necessarily singular.

The known restrictions on the ranks of PPT entangled
states considerably simplify the problem for two
qutrits. Every PPT entangled state in $3\otimes3$
has rank at least four~\cite{HSTT2003,Chen2008,
Chen2011a}. Consequently, a CES of dimension at
most three cannot support a PPT state. Since
the maximum dimension of a CES in $3\otimes3$
is four, it remains to determine which
four-dimensional CESs admit PPT entanglement.

An explicit example is the orthogonal complement of the
\emph{Pyramid} UPB of Bennett \emph{et
al.}~\cite{Bennett1999}. Its five orthogonal product vectors
are
\begin{equation}
\begin{aligned}
&\ket{\Psi_j}=\ket{\psi_j}\ot\ket{\psi_{2j\bmod 5}},\quad
j=0,\ldots,4, \quad {\rm With}\\
&\ket{\psi_j}=N\bigg[
\cos\!\left({2\pi j}/{5}\right)\ket0
+\sin\!\left({2\pi j}/{5}\right)\ket1
+\frac{\scriptstyle \sqrt{1+\sqrt5}}{\scriptstyle
2}\ket2\bigg],\\
&{\rm and}\quad  N=2/\sqrt{5+\sqrt5},
\end{aligned}
\label{eq:pyramidUPB}
\end{equation}
The orthogonal complement
$\Cc_{\mathrm{Pyr}}=\bigl(\spa\{\ket{\Psi_j}\}_{j=0}^4\bigr)
^\perp$ is a four-dimensional CES.
Its normalized projector
\begin{equation}
\rho_{\mathrm{Pyr}}=\frac14\left(I_9-
\sum_{j=0}^{4}\ket{\Psi_j}\!\bra{\Psi_j}\right)
\label{eq:pyramidstate}
\end{equation}
is PPT entangled~\cite{Bennett1999,DiVincenzo2003}. The
projector onto the UPB span is also PPT, since it is a sum
of orthogonal product projectors.
Solving the SDP in Eq.~\eqref{eq:muSDP} for
$\Cc_{\mathrm{Pyr}}$ gives $\mu(\Cc_{\mathrm{Pyr}})=0$. Here
PPT support is already
seen from the normalized projector.

The classification of rank-four PPT entangled states in
$3\otimes3$~\cite{Chen2011,Skowronek2011,
ChenDjokovic2012classes} allows us to turn this example into
a criterion that applies to the entire state space of any
four-dimensional CES.

\begin{proposition}[NPT property of four-dimensional CES in
$3\otimes3$]
\label{prop:generic-ces-npt}
Let $S\subset\C^3\ot\C^3$ be a four-dimensional CES. If $S$
is not related by invertible local operators to the
orthogonal complement of an orthogonal unextendible product
basis, then every density operator $\rho$ with
$\supp(\rho)\subseteq S$ is NPT. Equivalently, $\mu(S)<0$.
\end{proposition}
\begin{proof}
Suppose that a PPT state $\rho$ is supported on $S$. As $S$
is a CES, $\rho$ is entangled and $\rank\rho\leq4$. Every
two-qutrit PPT entangled state has rank at least
four~\cite{Chen2008}, so $\rank\rho=4$ and $\range\rho=S$.
The classification of rank-four two-qutrit PPT
states~\cite{Chen2011,Skowronek2011,ChenDjokovic2012classes}
gives an orthogonal UPB $\mathcal U$ and invertible local
operators $A,B$ for which
\begin{equation}
\rho=\alpha(A\ot B)(I_9-P_{\mathcal U})(A\ot B)^\dagger,
\qquad\alpha>0.
\label{eq:rank4-upb-form}
\end{equation}
Its range is therefore $(A\ot B)(\operatorname{span}\mathcal
U)^\perp$, contrary to the assumption on $S$. Therefore no
PPT state can be supported on $S$, and $\mu(S)<0$.
\end{proof}

The result gives a criterion for the entire subspace, rather
than for a particular density operator. Conversely, if a
four-dimensional CES is related to a UPB complement by
invertible local operators, transforming the corresponding
UPB-complement PPT state produces a PPT state on that CES.
Together with the rank restriction above, this characterizes
PPT support for all two-qutrit CESs.

The maximal CES introduced by Parthasarathy
~\cite{Parthasarathy2004} provides a useful contrast with
the Pyramid example. In $3\otimes3$ it has the form
\begin{equation}
\Cc_{\mathrm P}=\bigg\{\sum_{i,j=0}^{2}c_{ij}\ket{ij}:
\sum_{i+j=k}c_{ij}=0,\quad k=0,\ldots,4\bigg\}.
\label{eq:partha-CES}
\end{equation}
This four-dimensional CES also appears in equivalent
constructions by Bhat and
Johnston~\cite{Bhat2006,Johnston2013,Ritabrata2014}.
Sengupta \emph{et al.} showed that its normalized projector
is NPT and examined other positive operators supported on
it~\cite{Ritabrata2014}. In fact, $\Cc_{\mathrm P}$ cannot
support a PPT state. Its orthogonal complement contains the
product vectors
\begin{equation}
(\ket0+t\ket1+t^2\ket2)^{\otimes2},\qquad t\in\C,
\label{eq:partha-products}
\end{equation}
whereas the kernel of a rank-four two-qutrit PPT entangled
state contains only six projective product
vectors~\cite{Skowronek2011,ChenDjokovic2012classes}. Since
invertible local transformations preserve this distinction,
Proposition~\ref{prop:generic-ces-npt} gives
$\mu(\Cc_{\mathrm P})<0$. In contrast to the NPT normalized-
projector test, this establishes NPT for every density
operator supported on $\Cc_{\mathrm P}$.

\section{PPT support in higher-Schmidt-rank subspaces}
\label{sec:high-sr}
We now ask whether a PPT state can have a range that
excludes every vector of Schmidt rank one \emph{or} two.
This is stronger than having Schmidt number at least
three. A state of Schmidt number three may still contain
lower-Schmidt-rank vectors in its range: Schmidt number
constrains possible pure-state decompositions, whereas the
condition here concerns every vector in the supporting
subspace.

PPT states of Schmidt number greater than two have attracted
considerable interest. Krebs and Gachechiladze recently
constructed a $4\otimes5$ PPT state of Schmidt number three
by local extensions~\cite{KrebsGachechiladze2026}. As shown
below, no PPT state in this dimension can have a range
consisting entirely of vectors with Schmidt rank at least
three. Consequently, their example must contain a nonzero
vector of Schmidt rank at most two in its range.

In larger dimensions, PPT support is compatible with this
stronger requirement. Chen \emph{et al.} proved that the
tensor product of two states with completely entangled
ranges has no vector of Schmidt rank at most two in its
range~\cite{ChenYangTang2017}. For example, the state
 $\Omega=\rho_{\mathrm{Pyr}}^{\otimes2}$, regarded across
$A_1A_2|B_1B_2$, is PPT, has rank 16 in $9\ot9$, and
satisfies
\begin{equation}
\min_{0\neq\psi\in\range(\Omega)}\operatorname{SR}(\psi)\geq
3.
\label{eq:tensor-ppt-sr}
\end{equation}

The dimension bound in Eq.~\eqref{eq:rCESdim}, together with
restrictions on the ranks of PPT entangled states, rules out
PPT support on Schmidt-rank-three subspaces in several
low-dimensional systems.
\begin{proposition}[PPT support and minimum Schmidt rank
three]
\label{prop:sr3-ppt-rank}
No PPT state in $3\ot n$, $4\ot4$, or $4\ot5$ has a range
containing only vectors of Schmidt rank at least three. If a
PPT state in $4\ot6$ has this property, it must have rank
eight.
\end{proposition}
\begin{proof}
Let $d_A\leq d_B$ be the dimensions of the local supports of
such a state $\rho$, and let $r=\rank\rho$. The range
contains no product vector, so $\rho$ is PPT entangled. The
PPT rank bound and Eq.~\eqref{eq:rCESdim} imply
\begin{equation}
 d_B<r\leq(d_A-2)(d_B-2).
\label{eq:sr3-rank-bounds}
\end{equation}
These bounds exclude $d_A=3$ and $d_A=d_B=4$. In $4\ot5$,
the only remaining possibility has full local ranks and
$r=6$. By Theorem~43 of Ref.~\cite{ChenDjokovic2013PPT},
the range of a PPT entangled state of rank $d_B+1$,
when both local dimensions exceed three, necessarily
contains a product vector. A rank-six PPT state in
$4\otimes5$ therefore cannot satisfy our condition. In
$4\ot6$, the preceding results also exclude states with
reduced local supports, leaving only $r=7$ and $r=8$ as
possible ranks. The same theorem also rules out $r=7$.
\end{proof}

In $4\otimes6$, our results leave rank eight as the
only possibility for a PPT state whose range contains
no nonzero vector of Schmidt rank at most two.
Chen and {\DJ}okovi{\'c} have studied PPT states of
rank $d_A+d_B-2$, distinguishing between the good
and bad cases according to whether their kernels
contain finitely or infinitely many product
vectors~\cite{ChenDjokovic2013PPT}.
It remains unknown whether a rank-eight PPT state
can have a range in which every nonzero vector
has Schmidt rank at least three.
In $4\otimes5$, the PPT state of Schmidt number three
constructed by Krebs and Gachechiladze has product
vectors in its range~\cite{KrebsGachechiladze2026},
illustrating the distinction between Schmidt number
three and the stronger requirement considered here.

\section{Complementary completely entangled subspaces}
\label{sec:complementary}

We now turn to complementary CESs and investigate
whether both subspaces can simultaneously support PPT
entangled states. We first establish a general relation
between PPT states supported on orthogonal subspaces,
which holds even when the subspaces are not completely
entangled.

\begin{proposition}[PPT support on complementary subspaces]
\label{prop:complementary-mu}
Let $\Cc$ be a nonzero proper subspace of a bipartite
Hilbert space. If both $\Cc$ and $\Cc^\perp$ support PPT
states, then
\begin{equation}
\mu(\Cc)=\mu(\Cc^\perp)=0.
\label{eq:complementary-zero}
\end{equation}
In particular, $\mu(\Cc)>0$ implies $\mu(\Cc^\perp)<0$.
\end{proposition}
\begin{proof}
Let $\sigma$ be a PPT state supported on $\Cc^\perp$,
and let $\tau$ be any density operator supported on $\Cc$.
Their supports are orthogonal, so $\Tr(\tau\sigma)=0$.
Partial transposition preserves the trace of a product:
\begin{equation}
\Tr(\tau^{T_B}\sigma^{T_B})=\Tr(\tau\sigma)=0.
\end{equation}
Since $\sigma^{T_B}$ is positive and nonzero,
$\tau^{T_B}$ cannot be positive definite. Thus
$\mu(\Cc)\leq0$. The assumed PPT support on $\Cc$ gives
$\mu(\Cc)\geq0$, so equality holds. Interchanging the
subspaces proves the other equality. Finally, if
$\mu(\Cc)>0$, $\Cc^\perp$ cannot support a PPT state,
and hence $\mu(\Cc^\perp)<0$.
\end{proof}

Thus PPT states may occur on both sides of such a
decomposition, but neither subspace can support a state
with a positive-definite partial transpose. For CESs,
every such supported PPT state is also entangled.

In $3\otimes3$, two orthogonal completely entangled
subspaces cannot span the whole space, since each has
dimension at most four while the total dimension is nine.
The smallest bipartite system admitting complementary
completely entangled subspaces is $3\otimes4$, where each
can attain the maximal CES dimension of six. Examples of
these decompositions were given by Augusiak \emph{et
al.}~\cite{Augusiak2011CES} and
Skowronek~\cite{Skowronek2013}. We examine whether each
member of these pairs supports PPT states.

A first example comes from Augusiak \emph{et
al.}~\cite{Augusiak2011CES}:
\begin{equation}
\begin{aligned}
\Cc_A=\spa\{&\ket{01}-\ket{10},\ \ket{02}-\ket{20},\\
&\ket{03}-\ket{12}+\ket{21},\ \ket{13}-\ket{22},\\
&\ket{11}-\ket{23},\ \ket{00}-\ket{13}-\ket{21}\}.
\end{aligned}
\label{eq:augusiak-subspace}
\end{equation}
Both $\Cc_A$ and $\Cc_A^\perp$ are maximal completely
entangled subspaces. Their normalized projectors are NPT.
Numerical semidefinite optimization indicates an asymmetric
PPT structure,
\begin{equation}
\mu(\Cc_A)<0,\qquad \mu(\Cc_A^\perp)>0.
\label{eq:augusiak-mu}
\end{equation}
In particular, $\Cc_A^\perp$ supports PPT entanglement
even though its normalized projector is NPT, showing
that the normalized projector alone does not determine
PPT support. The values of $\mu(\Cc_A)$ and
$\mu(\Cc_A^\perp)$ were obtained through independent
SDP optimizations. Their signs are consistent with
Proposition~\ref{prop:complementary-mu}, according
to which $\mu(\Cc_A^\perp)>0$ necessarily implies
$\mu(\Cc_A)<0$.

A different pair arises in the one-parameter family studied
by
Skowronek~\cite{Skowronek2013}. For
$z\in\C\setminus\{0,1,-1\}$, let
$V_z=\spa\{\ket{v_i}:i=1,\ldots,6\}$ and
$V_z^\perp=\spa\{\ket{w_i}:i=1,\ldots,6\}$, where
\begin{equation}
\begin{aligned}
\ket{v_1}&=\ket{00}+\ket{11},
&\ket{v_2}&=\ket{10}+z\ket{21},\\
\ket{v_3}&=\ket{20}+z^{2}\ket{02},
&\ket{v_4}&=\ket{01}+z^{3}\ket{23},\\
\ket{v_5}&=\ket{12}+z^{4}\ket{03},
&\ket{v_6}&=\ket{22}+z^{5}\ket{13},
\end{aligned}
\label{eq:vbasis}
\end{equation}
\begin{equation}
\begin{aligned}
\ket{w_1}&=\ket{00}-\ket{11},
&\ket{w_2}&=\bar z\ket{10}-\ket{21},\\
\ket{w_3}&=\bar z^{\,2}\ket{20}-\ket{02},
&\ket{w_4}&=\bar z^{\,3}\ket{01}-\ket{23},\\
\ket{w_5}&=\bar z^{\,4}\ket{12}-\ket{03},
&\ket{w_6}&=\bar z^{\,5}\ket{22}-\ket{13}.
\end{aligned}
\label{eq:wbasis}
\end{equation}
Skowronek established that both subspaces are maximal and
completely entangled for these parameter
values~\cite{Skowronek2013}.

The structure of these subspaces can be understood by
arranging
the twelve computational-basis states on a $3\times4$ grid,
with $\ket{ij}$ corresponding to the site $(i,j)$.
Each vector $\ket{v_j}$ involves a distinct pair of sites:
\begin{equation}
\begin{aligned}
&(00,11),\quad (10,21),\quad (20,02),\\
&(01,23),\quad (12,03),\quad (22,13).
\end{aligned}
\label{eq:skow-pairs}
\end{equation}
These six pairs cover all twelve sites exactly once, forming
a perfect matching of the grid. The vectors $\ket{w_j}$
involve the same pairs, with each $\ket{w_j}$ orthogonal
to the corresponding $\ket{v_j}$.

For $q=|z|^2>0$, define the normalized states $\rho_v$
and $\rho_w$ by the following expressions, where $N_v$
and $N_w$ are the respective traces of the operators
in square brackets:
\begin{equation}
\begin{split}
\rho_v=\frac1{N_v}\bigl[{}&
\ket{v_1}\!\bra{v_1}
+q^{-1}\!\sum_{j=2}^3\ket{v_j}\!\bra{v_j}\\
&+q\!\sum_{j=4}^5\ket{v_j}\!\bra{v_j}
+\ket{v_6}\!\bra{v_6}\bigr],
\end{split}
\label{eq:skowR}
\end{equation}
\begin{equation}
\begin{split}
\rho_w=\frac1{N_w}\bigl[{}&
\ket{w_1}\!\bra{w_1}
+\ket{w_2}\!\bra{w_2}
+q^{-1}\ket{w_3}\!\bra{w_3}\\
&+q^{-4}\ket{w_4}\!\bra{w_4}
+q^{-5}\!\sum_{j=5}^6\ket{w_j}\!\bra{w_j}\bigr].
\end{split}
\label{eq:skowS}
\end{equation}

Both states are PPT. This can be
seen directly from the pairing structure above. Under
partial transposition, the off-diagonal operator
$\ket{ij}\bra{kl}$ becomes $\ket{il}\bra{kj}$.
In terms of the grid, this replaces the pair
$(ij,kl)$ by $(il,kj)$, as in the graphical description
of quantum grid states. For the six pairs in
Eq.~\eqref{eq:skow-pairs}, the transformation is
\begin{equation}
\begin{aligned}
(00,11)&\longmapsto(01,10),&
(10,21)&\longmapsto(11,20),\\
(20,02)&\longmapsto(22,00),&
(01,23)&\longmapsto(03,21),\\
(12,03)&\longmapsto(13,02),&
(22,13)&\longmapsto(23,12).
\end{aligned}
\label{eq:skow-ptpairs}
\end{equation}
Hence, the transformed pairs again cover all twelve
sites exactly once. Consequently, after a permutation
of the computational basis, $\rho_v^{T_B}$ and
$\rho_w^{T_B}$ decompose into six independent
$2\times2$ blocks.

For example, apart from the positive normalization
$N_v^{-1}$, the block of $\rho_v^{T_B}$ on
$\{\ket{01},\ket{10}\}$ is
\begin{equation}
\begin{pmatrix}
q&1\\
1&q^{-1}
\end{pmatrix},
\label{eq:skow-block}
\end{equation}
which is positive semidefinite with zero determinant.
The remaining blocks have the same property. Hence
\begin{equation}
\rho_v^{T_B}\geq0,\qquad
\rho_w^{T_B}\geq0,
\qquad
\rank\rho_v^{T_B}=\rank\rho_w^{T_B}=6.
\label{eq:skowPPT}
\end{equation}
Since the ranges are
completely entangled, both states are PPT entangled.
Proposition~\ref{prop:complementary-mu}
gives the following relations
\begin{equation}
\mu(V_z)=\mu(V_z^\perp)=0,
\qquad z\neq0,\pm1.
\label{eq:skowronek-mu}
\end{equation}

For $|z|=1$, the coefficients in
Eqs.~\eqref{eq:skowR} and \eqref{eq:skowS}
are equal and the six basis vectors in each subspace
have equal norm. In this case the corresponding
normalized orthogonal projectors are themselves PPT.
For $|z|\neq1$, the normalized projectors are NPT,
although $\rho_v$ and $\rho_w$ remain PPT.
The grid representation will also be useful in
understanding the $4\otimes4$ construction considered
below.

The preceding examples occur in $3\otimes4$. Motivated by
the pairing structure of the Skowronek construction and its
behavior under partial transposition, we now construct a
complementary pair in $4\otimes4$ for which both normalized
projectors are PPT, without requiring nonuniform weights.

Choose $\epsilon\in\{+1,-1\}$ and define
\begin{equation}
\begin{aligned}
\ket{u_1^\epsilon}&=\frac{\ket{00}-
\epsilon\ket{11}}{\sqrt2}, &
\ket{u_2^\epsilon}&=\frac{\ket{02}+\epsilon\ket{20}}{\sqrt2}
,\\
\ket{u_3^\epsilon}&=\frac{\ket{03}+\epsilon\ket{32}}{\sqrt2}
, &
\ket{u_4^\epsilon}&=\frac{\ket{12}+\epsilon\ket{33}}{\sqrt2}
,\\
\ket{u_5^\epsilon}&=\frac{\ket{01}+\epsilon\ket{23}}{\sqrt2}
, &
\ket{u_6^\epsilon}&=\frac{\ket{21}+\epsilon\ket{30}}{\sqrt2}
,\\
\ket{u_7^\epsilon}&=\frac{\ket{10}+\epsilon\ket{31}}{\sqrt2}
, &
\ket{u_8^\epsilon}&=\frac{\ket{13}+\epsilon\ket{22}}{\sqrt2}
,
\end{aligned}
\label{eq:ces44basis}
\end{equation}
and let $\Cc_\epsilon=\operatorname{span}\{
\ket{u_j^\epsilon}:j=1,\ldots,8\}$.
For either choice of sign the vectors have disjoint
supports, while $\ket{u_j^+}$ and $\ket{u_j^-}$ are
orthogonal
for every $j$.
Together, the sixteen vectors therefore form an orthonormal
basis, giving
$\Cc_- =\Cc_+^\perp$ and
$\dim\Cc_+=\dim\Cc_-=8$.
The subspaces here have dimension eight, whereas a maximal
CES in
$4\otimes4$ has dimension nine.

Next, we verify that neither subspace contains a product
vector. A general vector in $\Cc_\epsilon$ can be written as
\begin{equation}
\begin{split}
\ket{\phi_\epsilon}=\sqrt2\big(&a\ket{u_1^\epsilon}+b\ket{u_
2^\epsilon}
+c\ket{u_3^\epsilon}+d\ket{u_4^\epsilon}\\
&+e\ket{u_5^\epsilon}+f\ket{u_6^\epsilon}
+g\ket{u_7^\epsilon}+h\ket{u_8^\epsilon}\big),
\end{split}
\label{eq:ces44-general-vector}
\end{equation}
where $a,b,c,d,e,f,g,h\in\C$. The factor $\sqrt2$ cancels
the common normalization of the eight vectors in
Eq.~\eqref{eq:ces44basis}. Expressing $\ket{\phi_\epsilon}$
in the computational basis and grouping the terms by the
first subsystem gives
\begin{equation}
\begin{split}
\ket{\phi_\epsilon}={}&\ket0\ot(a\ket0+e\ket1+b\ket2+c\ket3)
\\
&+\ket1\ot(g\ket0-\epsilon a\ket1+d\ket2+h\ket3)\\
&+\ket2\ot(\epsilon b\ket0+f\ket1+\epsilon h\ket2+\epsilon
e\ket3)\\
&+\ket3\ot(\epsilon f\ket0+\epsilon g\ket1+\epsilon
c\ket2+\epsilon d\ket3).
\end{split}
\label{eq:ces44-row-expansion}
\end{equation}
For example, $\ket{u_1^\epsilon}$ contributes to the
coefficients of $\ket{00}$ and $\ket{11}$, with relative
sign $-\epsilon$; $\ket{u_2^\epsilon}$ contributes to the
coefficients of $\ket{02}$ and $\ket{20}$, with relative
sign $\epsilon$. The other six vectors are accounted for in
the same way. In the coefficient matrix $M_\epsilon$,
the first subsystem labels the rows and the second labels
the columns. The four parenthesized coefficient lists in
Eq.~\eqref{eq:ces44-row-expansion} therefore give
\begin{equation}
M_\epsilon=
\begin{pmatrix}
a&e&b&c\\
g&-\epsilon a&d&h\\
\epsilon b&f&\epsilon h&\epsilon e\\
\epsilon f&\epsilon g&\epsilon c&\epsilon d
\end{pmatrix}.
\label{eq:ces44matrix}
\end{equation}
The Schmidt rank of $\ket{\phi_\epsilon}$ is equal to $\rank
M_\epsilon$. A nonzero product vector would therefore
require $M_\epsilon$ to have rank one and hence all its
$2\times2$ minors to vanish. Suppose first that all eight
coefficients are nonzero. For a rank-one matrix, we have
\[
M_{11}M_{30}M_{22}M_{03}
=M_{00}M_{21}M_{13}M_{32},
\]
where indices start at zero. Substitution from
Eq.~\eqref{eq:ces44matrix} gives $-\epsilon afhc=\epsilon
afhc$,
which is impossible if all entries are nonzero.
If any of the eight coefficients is zero, the vanishing
of the $2\times2$ minors forces the remaining coefficients
to vanish as well. Hence neither subspace contains a
nonzero product vector, proving that both are CESs.

Let us examine whether the normalized projectors onto
these subspaces are PPT. They are given by
\begin{equation}
\rho_\epsilon=\frac18 P_{\Cc_\epsilon}
=\frac18\sum_{j=1}^8
\ket{u_j^\epsilon}\bra{u_j^\epsilon}.
\label{eq:ces44states}
\end{equation}
Under partial transposition, the off-diagonal terms
transform according to
$\ket{ij}\bra{kl}\mapsto\ket{il}\bra{kj}$.
For the states considered here, the resulting
off-diagonal entries couple the following pairs of
computational basis vectors,
\[
\begin{gathered}
(01,10),\ (00,22),\ (02,33),\ (13,32),\\
(03,21),\ (20,31),\ (11,30),\ (12,23).
\end{gathered}
\]
These eight pairs account for all sixteen computational
basis vectors without any overlap. Upon reordering the
basis accordingly, the partial transpose becomes a
direct sum of eight $2\times2$ matrices of the form
\begin{equation}
\frac1{16}
\begin{pmatrix}
1&s\\
s&1
\end{pmatrix},
\qquad s\in\{+1,-1\}.
\label{eq:ces44blocks}
\end{equation}
Their eigenvalues are $0$ and $1/8$, establishing
that both normalized projectors are PPT. Moreover,
\[
\operatorname{rank}(\rho_\epsilon)
=
\operatorname{rank}(\rho_\epsilon^{T_B})
=8.
\]
Since both subspaces are CESs, the corresponding
normalized projectors are PPT entangled.
Since we have already shown that their supports contain no
product
vectors,
$\rho_+$ and $\rho_-$ are PPT-entangled edge states. We also
find that
both reduced states are maximally mixed,
$\operatorname{Tr}_B\rho_\epsilon=
\operatorname{Tr}_A\rho_\epsilon=I_4/4$.

Since the two supports are orthogonal and both normalized
projectors are PPT, Proposition~\ref{prop:complementary-mu}
gives immediately
\begin{equation}
\mu(\Cc_+)=\mu(\Cc_-)=0.
\label{eq:ces44-mu}
\end{equation}

The partial-transpose blocks thus establish PPT positivity
of both complementary projectors.

\section{Conclusion}
\label{conclusion}
An NPT normalized projector does not necessarily
mean that a CES cannot support PPT entangled states.
We have studied this question by considering all
density operators supported on a given subspace.
For two qutrits, the classification of rank-four
PPT entangled states allows us to determine exactly
which CESs admit PPT entanglement, as illustrated by
the contrasting Pyramid and Parthasarathy examples.
We have also investigated whether PPT states can
have ranges containing only vectors of Schmidt rank
at least three. We show that this is impossible in
$3\otimes n$, $4\otimes4$, and $4\otimes5$, while
any such state in $4\otimes6$ must have rank eight.

Our study of complementary CESs shows that PPT
entanglement can exist on both sides of an orthogonal
decomposition of the Hilbert space. When this happens,
the partial transpose of every supported PPT state
must be singular. The Skowronek family in $3\otimes4$
provides examples where both subspaces support PPT
entangled states even though their normalized
projectors are NPT. In $4\otimes4$, we have constructed
a pair for which both normalized projectors are
themselves PPT entangled. This gives a construction
in which the entire Hilbert space is divided into
two orthogonal CESs, each supporting bound
entanglement. This gives us a direct way to construct two
PPT
entangled states with orthogonal completely entangled
ranges, without using a UPB complement.

It remains of interest to see whether all complementary CESs
 support PPT entanglement on at least one
side. The possibility of a rank-eight PPT state
with a Schmidt-rank-three range in $4\otimes6$ is another
question one can pursue. The relationship between these
complementary constructions and quantum grid
states~\cite{LockhartGuhneSeverini2018,Ghimire2023} may be
worth further investigation.

\section*{Acknowledgments}
The present work originated from discussions with Ritabrata
Sengupta more than a year ago. The authors thank him for
these discussions.

\noindent\textbf{Declaration on the use of AI:}
Following the completion of the main results, G.S. discussed
aspects of the work with OpenAI's ChatGPT (GPT-6), during
which possible connections between the $3\otimes4$
Skowronek, $4\otimes4$ constructions and quantum grid states
were identified. The authors take full responsibility for
the scientific content of the work.


\bibliographystyle{apsrev4-2}
\bibliography{ces}

\end{document}